\documentclass[aps,twocolumn,pre,showpacs,eqsecnum]{revtex4}

\usepackage{graphpap}
\usepackage[dvips]{graphicx}
\usepackage[dvips]{graphics}
\usepackage{color}

\begin{document}

\title{Tunable Coulomb blockade in nanostructured graphene}
 \author{C. Stampfer\footnote{Corresponding author, e-mail: stampfer@phys.ethz.ch}, J. G\"uttinger, F. Molitor, D. Graf, T. Ihn, and K. Ensslin}

\affiliation{Solid State Physics Laboratory, ETH Zurich, 8093 Zurich, Switzerland}

\date{ \today}
 
\begin{abstract}

We report on Coulomb blockade and Coulomb diamond measurements on an etched, tunable single-layer graphene quantum dot.
The device consisting of a graphene island connected via two narrow graphene constrictions is fully tunable by three lateral graphene gates. Coulomb blockade resonances are observed and from Coulomb diamond measurements a charging energy of~$\approx$~3.5~meV is extracted. For increasing temperatures we detect a peak broadening and a transmission increase of the nanostructured graphene barriers. 

\end{abstract}

\maketitle

\newpage

Graphene is a promising material~\cite{gei07,kat07} to investigate 
mesoscopic phenomena in two-dimensions (2d).
Unique electronic properties, such as massless carriers, electron-hole symmetry near the charge neutrality point, and weak spin-orbit coupling~\cite{min06} makes graphene interesting for high mobility electronics~\cite{che07,han07}, for tracing quantum electrodynamics in 2d solids, and for the realization of spin-qubits~\cite{tra07}. Whereas diffusive transport in graphene and the anomalous quantum Hall effect have been investigated intensively~\cite{nov05,zha05}, graphene quantum dots are still in their infancy from an experimental point of view~\cite{bun05}. This is mainly due to difficulties in creating tunable quantum dots in graphene because of the absence of an energy gap. Also phenomena related to
Klein tunneling make it hard to confine carriers laterally using electrostatic potentials~\cite{dom99,kat06}. 
Here we report on Coulomb blockade and Coulomb diamond measurements on an etched
graphene quantum dot tunable by graphene side gates~\cite{mol07}. 

The nanodevice, schematically shown in Fig.~1(a), has been fabricated from graphene, which has been extracted from bulk graphite by mechanical exfoliation onto 300~nm SiO$_2$ on n-Si substrate as described in Ref.~\cite{nov04}. 
Raman imaging~\cite{sta07a} is applied to verify the single-layer character of the investigated devices~\cite{fer06,gupta,dav07a}. 
90~nm PMMA (positive e-beam resist) is then spun onto the samples and electron-beam (e-beam) lithography is
used to pattern the etch mask for the graphene devices. Reactive ion etching (RIE) based on an Ar/O$_2$ (9:1) plasma is introduced to etch away unprotected graphene. 
A scanning force microscope (SFM) image of the etched graphene structure after removing the residual PMMA is shown in Fig.~1(b). 
Finally, the graphene device is contacted by e-beam patterned 2~nm Ti and 50~nm Au electrodes as shown in Fig.~1(c).
A Raman spectrum recorded on the final device taken at the location of the graphene island is plotted in Fig.~1(e). It is an unambiguous fingerprint of single-layer graphene with a line width of the 2D line of approx. 33~cm$^{-1}$~\cite{fer06,gupta,dav07a}. The elevated background originates from the nearby metal electrodes and the significant D line is due to edges within the area of the laser spot size of $\approx$~400~nm.
In addition to Raman spectroscopy, the SFM step height of $\approx 0.5$~nm, shown in Fig.~1(d), proves also the single-layer character of the graphene flake and shows that the RIE etching does not attack the SiO$_2$. 

The fabricated device consists of two $\approx$ 50~nm
narrow graphene constrictions
connecting source (S) and drain (D) electrodes to a graphene island with an area $A$~$\approx~0.06~\mu m^2$. The twoelectrostatically the two barriers and the island, respectively. For assignment of the gate electrodes see Fig.~1(a). All three graphene side gates have been patterned
 closer than 100~nm to the 
active graphene
 \begin{figure}[hbt]\centering
\includegraphics[draft=false,keepaspectratio=true,clip,%
                   width=0.96\linewidth]%
                   {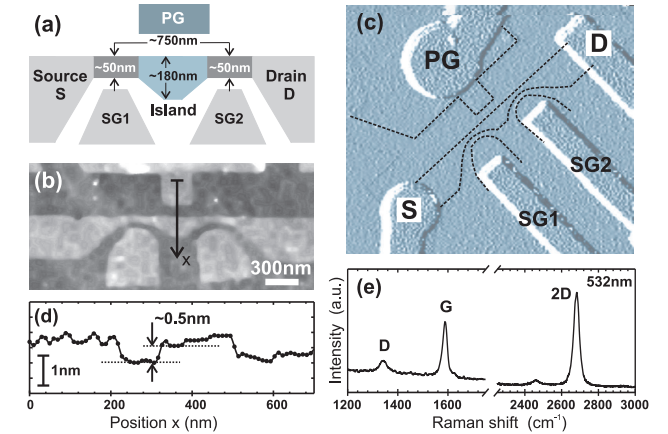}                   
\caption[FIG1]{(color online)
Nanostructured graphene quantum dot device. (a) Schematic illustration of the tunable graphene quantum dot. (b) Scanning force microscope (SFM) image of the investigated graphene device after RIE etching and (c) after contacting the graphene structure. The minimum feature size is approx.~50~nm. 
The dashed lines indicate the outline of the graphene areas.
(d) shows a SFM cross-section along a path $x$ [marked in (b)] averaged over $\approx$~40~nm perpendicular to the path proving the selective etch process. (e) Confocal Raman spectra recorded on the final device at the graphene island with a spot size of approx.~400~nm, clearly proving the single-layer character of the investigated device. For more information on the D, G and 2D (also called D') line please refer to Ref.~\cite{dav07a}.} 
\label{trdansport}
\end{figure}
 regions, as shown in Figs.~1(b,c). The additional back gate (BG) is used to adjust
 the overall Fermi energy.

Transport measurements have been performed in a variable temperature He cryostat at a base temperature of $ \approx$~1.7~K. Before the cool-down the sample has been baked in vacuum at 135$^{\circ}$C for 12~h. We have measured the two-terminal conductance through the dot by applying a small (symmetric) DC or AC bias voltage $V_{bias}$, and measuring the current through the dot with a resolution better than 20~fA. 
At high bias (e.g., $V_{bias}=100$~mV, not shown), the (back) gate characteristics clearly reveal the charge neutrality point of the graphene material. Such measurements are used to adjust the range of the back gate voltage. In the following we
kept the back gate fixed close to the overall charge neutrality point at $V_{BG}=-6$~V, where transport can be pinched off by the two side gates $V_{SG1}$ and $V_{SG2}$.
At small bias ($V_{bias}<200$~$\mu$V) transport is dominated (i) by the two narrow junctions, where strong transmission modulations and gap effects appear, and (ii) by Coulomb blockade due to charging of the graphene island.
Both effects can be seen in Fig.~2, where the source-drain current is plotted as a function of the two barrier gate voltages $V_{SG1}$ and $V_{SG2}$ for constant $V_{bias}=200$~$\mu$V.
The large scale horizontal and vertical current modulations can be attributed to either
one or the other narrow graphene constriction, being tuned (almost) independently from each other. On top we observe Coulomb resonances which are associated with charging of the graphene island and, thus, tuned by both side gate potentials $V_{SG1}$ and $V_{SG2}$ (diagonal lines).
\begin{figure}[t]
\centering
\includegraphics[draft=false,keepaspectratio=true,clip,%
                   width=0.95\linewidth]%
                   {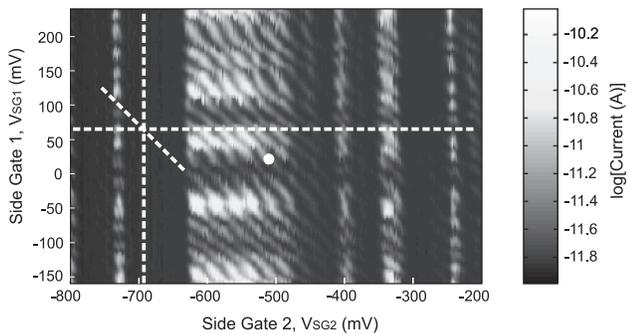}                   
\caption[FIG1]{
Source-drain current as a function of the two barrier gate voltages $V_{SG1}$ and $V_{SG2}$ for constant bias, $V_{bias}=200 \mu V$. The dashed lines indicate transmission modulations and oscillations attributed to the graphene constrictions (horizontal and vertical lines) and to the island (diagonal line). Measurements are preformed at $V_{BG}=-6$~V and $V_{PG}=0$~V.} 
\label{trdansport}
\end{figure}

By sweeping $V_{SG1}$ and $V_{SG2}$ to a regime where the background current is significantly suppressed (see white point in Fig.~2), the plunger gate $V_{PG}$ can be used to trace Coulomb resonances as shown in Fig.~3(a). 
In this configuration of gate voltages the peak positions were stable in more than 10 consecutive plunger gate sweeps. 
\begin{figure}[t]\centering
\includegraphics[draft=false,keepaspectratio=true,clip,%
                   width=0.85\linewidth]%
                   {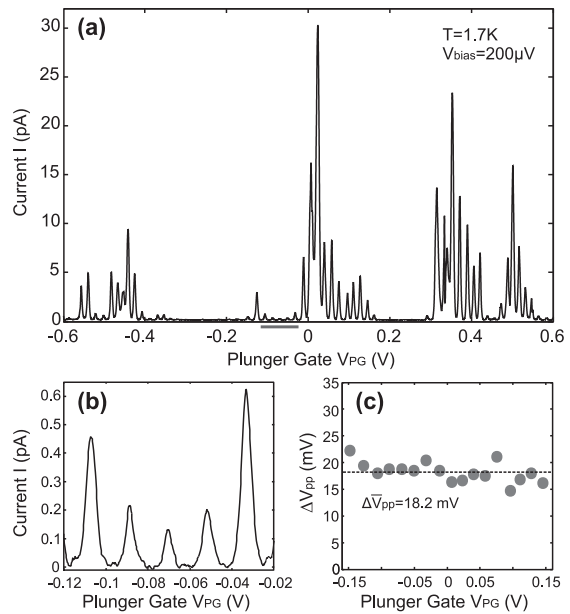}                   
\caption[FIG1]{
Source-drain current through the graphene nanostructure as function of the plunger gate voltage $V_{PG}$. (a) clear Coulomb resonances are observed on top and next to large scale conductance modulations. (b) shows a marked close-up of (a) and in (c) the peak spacing is plotted for 18 consecutive peaks. Measurements are preformed in the dot configuration: $V_{BG}=-6$~V, $V_{SG1}=25$~mV, and $V_{SG2}=-510$~mV.}
\end{figure}
\begin{figure}[t]
\centering
\includegraphics[draft=false,keepaspectratio=true,clip,%
                   width=1\linewidth]%
                   {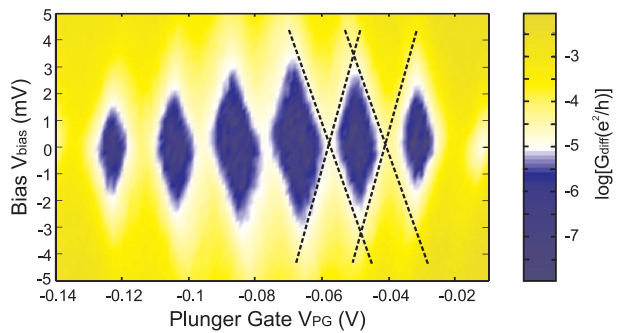}                   
\caption[FIG1]{
(color online) 
Coulomb diamonds in differential conductance $G_{diff}$, represented in a logarithmic color scale plot (dark regions represent low conductance). A DC bias $V_{bias}$ with a small AC modulation (50~$\mu$V) is applied symmetrically across the dot and the current through the dot is measured. Differential conductance has been directly measured by a Lock-in amplifier. The charging energy is estimated to be $\approx 3.6$~meV from this measurements. Measurements are preformed in the dot configuration: $V_{BG}=-6$~V, $V_{SG1}=25$~mV, and $V_{SG2}=-510$~mV.} 
\label{trdansport}
\end{figure}
Among regions where transport is completely pinched off by the narrow constrictions, large
scale conductance modulations in the barriers are observed.
Nearby and on top of these large features clear Coulomb peaks are measured (see e.g. Fig.~3(b), which is a close-up of Fig.~3(a)).
The period of the Coulomb oscillations measured over 18 consecutive peaks is $\Delta \overline{V}_{pp}~\approx~18.2$~mV, as shown in Fig.~3(c). There are no systematic peak spacing fluctuations, and the observed deviations might be influenced by the underlying transmission modulation in both narrow constrictions.
However, the distribution of the nearest-neighbor spacing of the Coulomb oscillations is significantly larger than 
expected for purely metallic single-electron transistors~\cite{fur00}.

Coulomb diamond measurements~\cite{kou97}, i.e., measurements of the differential conductance ($G_{diff}$) as function of symmetric bias voltage $V_{bias}$ and plunger gate voltage $V_{PG}$ are shown in Fig.~4.
The elevated background at the left and right side is due to barrier dependent conductance modulations as shown in Fig.~3(b). 
Please note that within the swept plunger gate voltage range no charge rearrangements have been observed.
From the extent of the diamonds in bias direction we estimate the charging energy of the graphene dot to be $E_C~\approx~3.5$~meV. This charging energy corresponds to a capacitance of the dot $C=e^2/E_C~\approx~45.8$~aF. The lever arm of the plunger gate is $\alpha_{PG}=C_{PG}/C~\approx~0.19$. 
The electrostatic coupling of all other lateral gates were determined~\cite{kou97} to be $C_{SG1}~\approx~3.9$~aF, $C_{SG2}~\approx~5.9$~aF, and $C_{PG}~\approx~8.7$~aF. The extracted back gate capacitance $C_{BG}~\approx~18$~aF is slightly higher than the purely geometrical parallel plate 
capacitance of the graphene island $C=\epsilon_0 \epsilon A/d~\approx~7.4~aF$. This is not surprising since $\sqrt{A} \approx d$, where $A$ is the area of the graphene island and $d$ the gate oxide thickness.
A screened Hartree approx. can easily account for a factor 2~\cite{ihn04}.

\begin{figure}[t]
\centering
\includegraphics[draft=false,keepaspectratio=true,clip,%
                   width=0.86\linewidth]%
                   {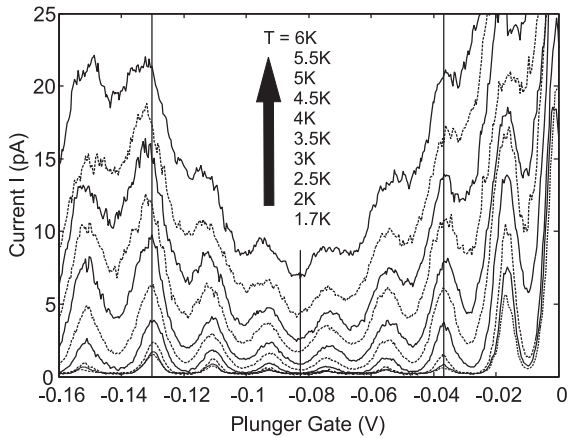}                   
\caption[FIG1]{
Source drain current as function of the plunger gate voltage $V_{PG}$ for different bath temperatures. Note that, the plunger gate sweep includes the region shown in Fig.~3b. 
 Measurements are preformed in the dot configuration: $V_{BG}=-6$~V, $V_{SG1}=25$~mV, and $V_{SG2}=-510$~mV. The different bath temperatures are indicated.} 
\label{trdansport}
\end{figure}

We further explored the temperature dependence of the Coulomb oscillations, as shown in Fig.~5 for a temperature range from $T \approx 1.7$ to $\approx 6$~K. We found (i)
a strong increase of the barrier dependent background and (ii) a broadening and initial increase of the Coulomb peak heights. The overall temperature dependent current increases significantly and depends on the applied gate voltage. Thus, the transmission modulations in the barriers, which are attributed to gap effects~\cite{han07} are highly temperature dependent, as seen by the elevated current background (as indicated along vertical lines in Fig.~5). At a bath temperature $T \approx 1.7$~K we obtain a carrier temperature of $T_{e} \approx 2.2$~K by fitting the Coulomb peaks in the multi-level transport regime~\cite{bee90}.  

In conclusion, we have fabricated a tunable graphene quantum dot based on RIE patterned graphene with graphene lateral gates. Its functionality was demonstrated by observing clear and reproducible Coulomb
resonances. From the Coulomb diamond measurements it was estimated that the charging energy of the graphene dot is $\approx$~3.5~meV, which is compatible with its lithographic dimensions. Excited states are not observed in Coulomb diamond measurements,
which is due to the rather large graphene dot size and the elevated carrier temperature.
However, these results open the way to more detailed studies of graphene quantum dots at even lower temperatures, where the transport through single quantum states is anticipated.

{Acknowledgment ---}
The authors wish to thank R.~Leturcq, P.~Studerus, C.~Barengo, P.~Strasser, F.~Libisch and K.~S.~Novoselov for helpful discussions. Support by the ETH FIRST Lab and financial support by the Swiss National Science Foundation and NCCR nanoscience are gratefully acknowledged.

\end{document}